\documentclass[%
reprint,
superscriptaddress,
amsmath,amssymb,
aps,
prl,
longbibliography
]{revtex4-2}

\usepackage{graphicx}
\usepackage{dcolumn}
\usepackage{siunitx}
\usepackage{bm}
\usepackage{upgreek}
\usepackage{float}

\usepackage{xcolor}
\usepackage[colorlinks = true,
            linkcolor = blue,
            urlcolor  = blue,
            citecolor = blue,
            anchorcolor = blue]{hyperref}
\usepackage{color,soul} 

\graphicspath{{Images/}}

\begin{document}
\preprint{APS/123-QED}

\title{Probing the Degree of Coherence through the Full 1D to 3D crossover}

\author{R. Shah}
\author{T. J. Barrett}
\affiliation{Department of Physics and Astronomy, University of Sussex, Brighton BN1~9QH, UK}%
\author{A. Colcelli}
\affiliation{SISSA and INFN, Sezione di Trieste, Via Bonomea 265, I-34136 Trieste, Italy}
\author{F. Oru\v{c}evi\'{c}}
\affiliation{Department of Physics and Astronomy, University of Sussex, Brighton BN1~9QH, UK}
\author{A. Trombettoni}
\affiliation{Department of Physics, University of Trieste, Strada Costiera 11, I-34151 Trieste, Italy}
\affiliation{SISSA and INFN, Sezione di Trieste, Via Bonomea 265, I-34136 Trieste, Italy}
\author{P. Kr\"{u}ger}
\affiliation{Department of Physics and Astronomy, University of Sussex, Brighton BN1~9QH, UK}
\affiliation{Physikalisch-Technische Bundesanstalt, 10587 Berlin, Germany}
\date{\today}

\begin{abstract}
We experimentally study a gas of quantum degenerate $^{87}$Rb atoms throughout the full dimensional crossover, from a one-dimensional (1D) system exhibiting phase fluctuations consistent with 1D theory to a three-dimensional (3D) phase-coherent system, thereby smoothly interpolating between these distinct, well-understood regimes. Using a hybrid trapping architecture combining an atom chip with a printed circuit board, we continuously adjust the system's dimensionality over a wide range while measuring the phase fluctuations through the power spectrum of density ripples in time-of-flight expansion. Our measurements confirm that the chemical potential $\mu$ controls the departure of the system from 3D and that the fluctuations are dependent on both $\mu$ and the temperature $T$. Through a rigorous study we quantitatively observe how inside the crossover the dependence on $T$ gradually disappears as the system becomes 3D. Throughout the entire crossover the fluctuations are shown to be determined by the relative occupation of 1D axial collective excitations.
\end{abstract}

\maketitle{}

The dimensionality of a system can have dramatic effects on its properties, giving rise to a plethora of interesting behavior. The nature of superfluid and superconducting phase transitions is well known to be radically different in systems of one, two, or three dimensions. The Mermin-Wagner-Hohenberg theorem \cite{Mermin1966,Hohenberg_1967} dictates that at finite temperature more than two dimensions are required for true long-range order. The transition in two dimensions is governed by a Kosterlitz-Thouless mechanism \cite{hadzibabic2006berezinskii} of topological origin, and for three dimensions it is the paradigmatic example of symmetry breaking that is qualitatively well-described by mean-field theories. In one dimension no such transition exists, but due to the enhanced role of both quantum and thermal fluctuations there is a richer set of physical regimes than in two or three dimensions \cite{Kheruntsyen_PairCorrelations1DGas,petrov2000regimes, bouchoule2011atom}. 

The stark contrast of transition phenomena makes the study of a system that lies between two distinct dimensions -- in a \textit{dimensional crossover} -- of great fundamental interest, as well as offering the potential for practical applications. A typical example is provided by layered superconductors, either naturally occurring \cite{Gamble1970} or artificially controlled \cite{Ruggiero1980}, presenting instances of the 2D to 3D crossover. While this has been extensively studied, producing superconducting samples in the 1D to 3D crossover is technologically more challenging, but remains a subject of intense research, with the ultimate goal to realize new high-temperature superconductors \cite{Perali1996_HighTcq1Dsuperconductors,Biancoli1997_HighTcq1Dsuperconductors, Shanenko2006_HighTcq1Dsuperconductors, Mazziotti2017_HighTcq1Dsuperconductors, Saravia2020_HighTcq1Dsuperconductors}. Alternatively, the 1D to 3D crossover could be partially accessed with superfluid $^4$He inside carbon nanotubes and nanopores, with the 1D regime being reached when the transverse size becomes on the order of a few angstroms \cite{DelMaestro,Hauser2017}--which is currently very difficult to obtain \cite{Li2017_CNT}.

Conversely, ultracold atom experiments are naturally suited to study the 1D to 3D crossover, where the external trapping geometry can be flexibly tuned to constrain atomic degrees of freedom, providing the means to effectively manipulate the dimensionality. Examples include single magnetically trapped systems \cite{Gorlitz2001_LowD_BECs}, or arrays of systems with tunable coupling in optical lattices \cite{Stoferle2004_1D3Dlattices}. In particular, purely 1D systems can now be routinely formed by employing the extremely tight traps generated by \textit{atom chips} \cite{Folman_2000, AtomChips2011_Ch2,kruger2010weakly}. A major difference to 3D systems is the presence of both density \cite{Armijo_2011,Jacqmin2011_StronglyInteracting1D} and phase fluctuations, the latter having been studied by several experiments in various limited regimes \cite{Dettmer_2001,Hellweg_2001,Richard_2003,Manz_2010,Gallucci2012_PhaseyThings,Schemmer_2018}. However, a comprehensive experimental mapping of the phase fluctuations in the entire crossover remains elusive, since few experiments have access to the necessary tunability of trapping geometry, atom number, and temperature.

\begin{figure*}[!ht]
\centering
\includegraphics[width=\textwidth]{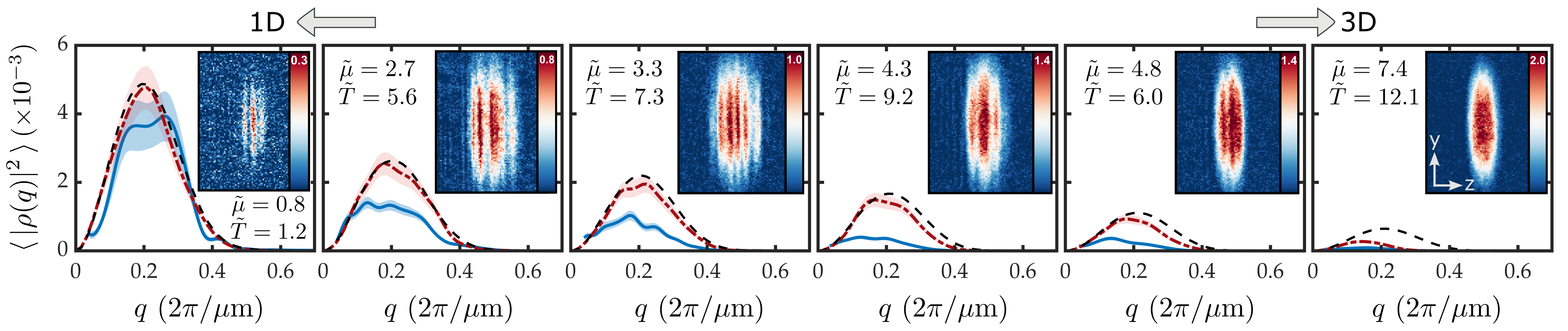}
\caption{\label{fig:Figure_1} Measured density ripples power spectra (solid blue lines) for a number of traps through the 1D to 3D dimensional crossover, ordered by $\tilde{\mu} = \mu/\hbar\omega_\perp$ and $\tilde{T} = k_\mathrm{B}T/\hbar\omega_\perp$, with trapping frequencies from left to right \mbox{$\omega_\perp = 2\pi\times$(1235, 1380, 1005, 800, 740, 395) Hz} and \mbox{$\omega_z = 2\pi\times$(16.1, 17.4, 16.9, 20.8, 23.2, 33.7) Hz}. For comparison the theoretical predictions are shown, both with (red dash-dotted lines) and without (black dashed lines) the effect of interactions during time of flight. Finite optical imaging resolution has been accounted for by convolution with a Gaussian of width $\sigma_\mathrm{psf} = \SI{4}{\micro\metre}$ (see Supplementary Material \cite{Imaging}).  Shaded bands represent the statistical uncertainty as 2 standard deviations from the mean, and are obtained via bootstrapping. Insets each depict a representative example of an experimental OD image, showing an individual realization from the corresponding dataset. 
}
\end{figure*}

Here, we map out the full 1D to 3D crossover by characterizing the phase fluctuations in individual degenerate Bose gases over a wide parameter space from a chemically 1D system (i.e., with \mbox{$\mu\lesssim \hbar\omega_\perp$} where $\mu$ is the zero temperature chemical potential and $\omega_\perp$ is the transverse trapping frequency), through to the 3D regime with \mbox{$\mu\gg\hbar\omega_\perp$}, where the fluctuations smoothly disappear. Using a combination of an atom chip and printed circuit board (PCB), we are able to control independently the axial and radial confinement, allowing measurement of the phase fluctuations across a wide range of external trap aspect ratios, in addition to studying a range of temperatures and atom numbers. The in-trap phase fluctuations are studied by observing the power spectrum of the density ripples which form as the atomic cloud expands during time of flight. It is well understood that when both \mbox{$\mu$, $k_\mathrm{B}T \ll \hbar\omega_\perp$} the system is 1D, and in this regime our experimental results agree well with 1D theoretical predictions \cite{Imambekov_2009}. For larger \mbox{$\mu/\hbar\omega_\perp$} we observe an increasing divergence from the 1D theory (Fig.\ \ref{fig:Figure_1}). The common expectation derived from a number of previous studies is that both \mbox{$k_\mathrm{B}T/\hbar\omega_\perp$} and \mbox{$\mu/\hbar\omega_\perp$} are relevant in the crossover \cite{khawaja_2003s, vanAmerongen2008_YYonAtomChip, Hadzibabic_2008, kruger2010weakly, Armijo_2011, Gallucci2012_PhaseyThings, RuGway_2013, Moller_2021}. Here we provide a detailed quantitative study of the interplay between these two parameters that has so far been missing. Access to the dimensional crossover regime is driven most strongly by the reduced chemical potential \mbox{$\tilde{\mu} = \mu/\hbar\omega_\perp$}. Then, inside the crossover the fluctuations become increasingly dependent on the reduced temperature \mbox{$\tilde{T}=k_\mathrm{B}T/\hbar\omega_\perp$} as the 1D regime is approached, but become less sensitive to $\tilde{T}$ towards three dimensions. We show that throughout the crossover the strength of the density ripples is determined by the relative occupation of low-energy axial excitations, and not by either \mbox{$\tilde{\mu}$} or $\tilde{T}$ alone. 

Our experiment uses $^{87}$Rb atoms prepared in the \mbox{$\left|F=2,m_F=2\right\rangle$} magnetic substate which are loaded into a cylindrically symmetric Ioffe-Pritchard style wire \emph{H} trap. Transverse confinement ($x$-$y$ plane) is realized by a current-carrying wire on the atom chip together with an external bias field. Independent axial confinement (along $z$) is provided by two parallel wires on the PCB below the atom chip. We vary the current in the trapping wires to create a number of potentials with varying aspect ratio \mbox{$\kappa = \omega_{\perp}/\omega_{z}$}, with trapping frequencies in the range $\omega_{\perp} / 2 \pi = \SIrange[range-phrase =-,range-units = brackets]{570}{1380}{Hz}$ transversely, and $\omega_{z}/2\pi=\SIrange[range-phrase =-,range-units = brackets]{15}{34}{Hz}$ axially. 

After loading precooled atoms (\SI{\sim 10}{\micro\kelvin}) into the \emph{H} trap, condensates of approximately $10^5$ atoms are produced after \mbox{$\SI{1.5}{\second}$} of radio-frequency (rf) evaporation, and are then held in a \mbox{$\SI{12}{\kilo\hertz}$} rf shield \cite{Beijerinck_2000} for a further \mbox{$\SI{150}{\milli\second}$} to ensure thermal equilibrium (equating to several tens to hundreds of collisions per particle for all traps considered \cite{Davis1995_EvapCool}). An additional adjustable hold time of up to \mbox{$\SI{700}{\milli\second}$} is applied to allow for controllable losses through background collisions, which varies the final atom number in the range $N=\SIrange[scientific-notation = fixed, range-units = brackets, fixed-exponent = 4, range-phrase =-]{0.5e4}{10e4}{}$. By adjusting the final rf evaporative cooling frequency the temperature of the sample is set in the range $T = \SIrange[range-phrase =-,range-units = brackets]{70}{540}{\nano\kelvin}$, corresponding to $T/T_c = \numrange[range-phrase =-,range-units = brackets]{0.5}{0.8}$. Optical density (OD) images are acquired via standard absorption imaging \cite{Smith2011_ImagingOnChip} with a probe beam along the $x$ direction after a time-of-flight $t_\mathrm{tof} = \SI{34}{\milli\second}$. To suppress undesirable diffraction fringes in the OD images, we ensure optimal focusing of the imaging objective following the technique described in \cite{Putra_2014} (see Supplemental Material \cite{Imaging}). The insets in Fig.~\ref{fig:Figure_1} show a set of typical OD images exhibiting the density ripples of varying strength dependent on dimensionality.

To quantitatively analyze the spatial frequency content of the images, we calculate the power spectrum of the density ripples using the following steps. First, several hundred OD images of clouds under a chosen set of experimental conditions are acquired and postselected such that the standard deviation in atom number and temperature is approximately $5$\% of the respective value of the set. The thermal component of the gas is then fitted to each image to obtain the temperature \cite{Ensher1996_ThermalFit,Gerbier2004_ThermalFit,Szczepkowski2009_ThermalFit}, and is then removed before further analysis. Next, each column density image is integrated along the remaining transverse direction of the cloud (i.e. along $y$) to obtain the axial 1D line density $n_i(z)$, which is then subtracted from the mean of the set $\langle n(z) \rangle$, leaving only the residual line density $\delta n_i(z) = n_i(z) - \langle n(z) \rangle$ for each individual shot $i$. We then Fourier-transform these density residuals
\begin{equation}
\label{Fourier}
    \delta \tilde{n}_i(q) = \int \delta n_i(z)\, e^{-iqz} \; \textrm{d}z,
\end{equation}
where $q$ is the angular spatial frequency, and use $\delta \tilde{n}_i(q)$ to calculate a dimensionless power spectrum for each individual realization, normalized by atom number
\begin{equation}
\label{modulus}
    |\rho_i(q)|^2 = \frac{1}{\tau N_i^2}\left| \delta \tilde{n}_i(q) \right|^2,
\end{equation}
where $\tau = \omega_z t_\mathrm{tof}$. Note that dividing by $\tau N_{i}^2$ removes dependencies on total atom number and system length (see Supplemental Material \cite{Imaging}). Finally, we compute the mean power spectrum for the ensemble $\langle|\rho(q)|^2\rangle$, providing a single spectrum for each set of experimental conditions.

Figure \ref{fig:Figure_1} shows examples of typical experimental data, together with the corresponding density ripples spectra. Parameters ($N$, $T$, $\omega_\perp$, $\omega_z$) are varied to move from a 3D condensate with no visible density ripples to a deeply quasicondensate regime with strong density ripples. The theoretical predictions are generated using a stochastic model for the in-trap phase distribution that reproduces Bogoliubov results \cite{Stimming_2010}, and has been applied successfully in the 1D  \cite{Betz_2011, Langen_2013}, 2D \cite{Mazets_2012}, and elongated 3D (if the phase varies only axially) regimes \cite{Petrov_2001}. Many such realizations of a \mbox{one-dimensional} phase $\phi(z)$ are generated and imprinted onto the zero-temperature ground state wave function \mbox{$\psi(\mathbf{r}) = \sqrt{n(\mathbf{r})}e^{i\phi(z)}$}, constituting an ensemble of initial states (see Supplemental Material \cite{Imaging}). To obtain the density profiles after time of flight in the absence of interactions during expansion, we numerically propagate the initial states using the free Schr\"{o}dinger equation. The density ripple power spectra are then extracted in the same manner as for the experimental data. In order to account for modifications due to the effects of mean-field interactions, we additionally propagate the same initial states using the full 3D Gross-Pitaevskii equation (GPE), with the corresponding spectra also shown in Fig.~\ref{fig:Figure_1} for comparison. Such simulations are computationally demanding, and so an analytic hydrodynamic scaling approach has been previously used for clouds with minimal axial expansion (although with a discrepancy between the simulated and measured size of the fluctuations) \cite{Hellweg_2001}. Alternatively, experiments can be restricted to the simpler case in which it is valid to neglect interactions during expansion, limiting studies to high aspect ratios of typically $\gtrsim$100 \cite{Manz_2010,Schemmer2018_Thermometry}. We note that in this case a simple analytic expression for the power spectrum is available \cite{Imambekov_2009}, and can be extended to account for inhomogeneous density profiles using a local density approximation \cite{Schemmer_2018}. Here, using the combination of a graphics processing unit together with the algorithm developed in Ref.~\cite{Deuar2016} we are able to directly simulate the full TOF expansion, including mean-field interactions, for several thousands of realizations.

\begin{figure}[t!]
 \centering
\includegraphics[width=0.5\textwidth]{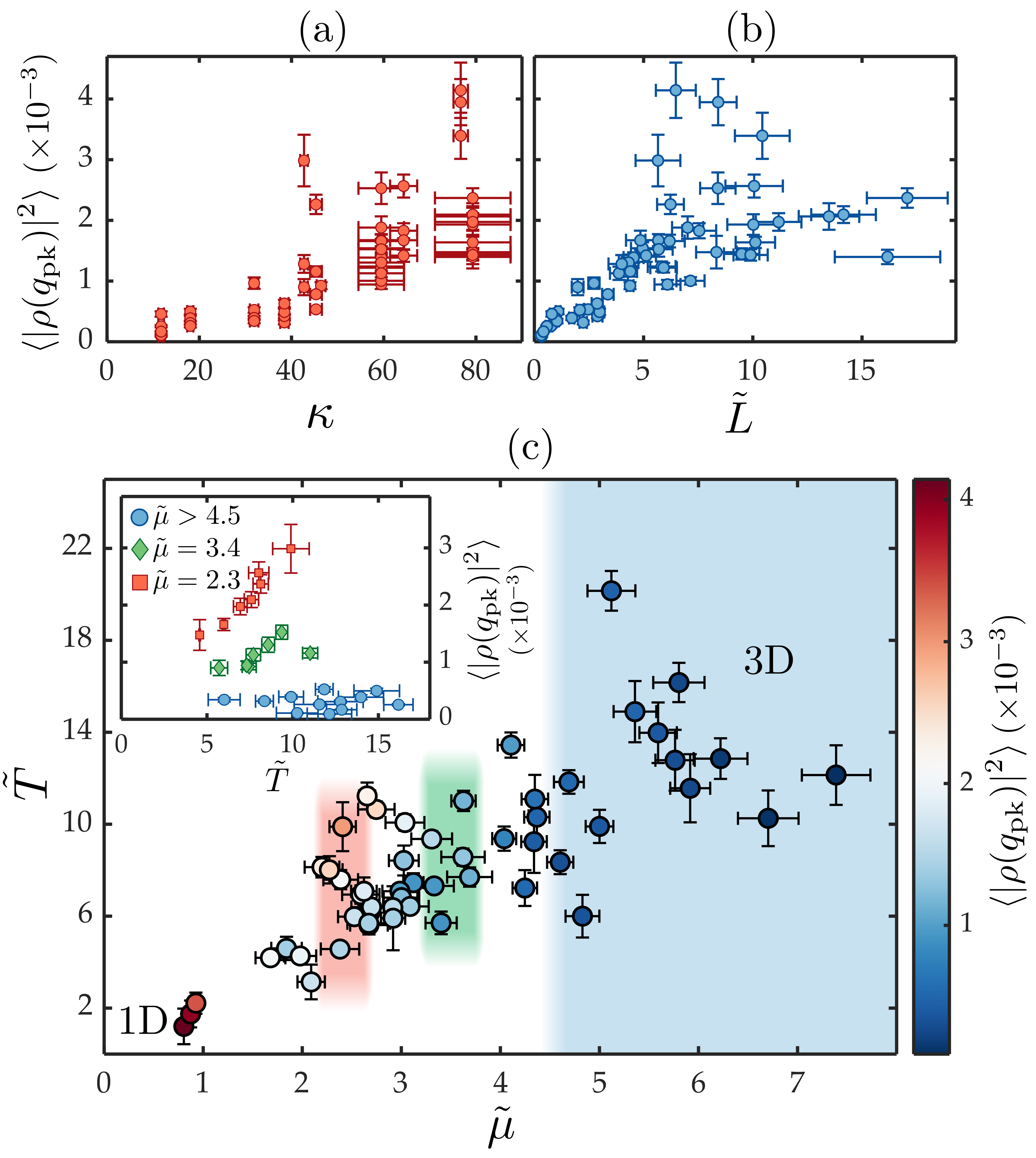}
\caption{\label{fig:Figure_2} Dependence of the amplitude of the phase fluctuations on (a) the aspect ratio of the trapping potential $\kappa$, and (b) the ratio of system length to phase coherence length $\tilde{L}$. (c) Visualisation of the parameter space explored in terms of $\tilde{T}$ and $\tilde{\mu}$. The color scale represents the peak amplitude of the power spectrum $\langle|\rho(q_\mathrm{pk})|^2\rangle$ for the measured data. Shaded bands indicate the data used for the line graphs in the inset. (Inset) Peak amplitude versus $\tilde{T}$ for (blue circles) $\tilde{\mu} > 4.5$, (green diamonds) $\tilde{\mu} \sim 3.4$, and (red squares) $\tilde{\mu} \sim 2.3$. All error bars show the statistical uncertainty as two standard deviations from the mean, and are obtained by bootstrapping.}
\end{figure}

At the 1D side of the crossover ($\tilde{\mu} \lesssim 1$) the quasicondensates display large phase fluctuations and both theoretical models show excellent agreement with each other and with the experimental data in Fig.~\ref{fig:Figure_1}, justifying a noninteracting expansion in this regime. The fluctuations at the 3D end of the crossover are suppressed \cite{Perrin2012, Richard_2003}, as expected for a true 3D BEC with long-range order, with full phase coherence across the ensemble even at finite temperature \cite{andrews1997observation,hagley1999measurement,stenger1999bragg,bloch2000measurement}. Between these two well-understood distinct regimes, we measure significant fluctuations even when \mbox{$\tilde{\mu}, \tilde{T} \ll 1$} is not satisfied, consistent with previous results \cite{Dettmer_2001, Hellweg_2001} where it was found that such systems can acquire some 1D characteristics (typical of a quasicondensate) \cite{Petrov_2001} when the phase coherence length is smaller than the extent of the sample \cite{Hellweg_2003}. Neither theoretical model fully accounts for the observed reduction in amplitude -- however the inclusion of interactions captures well the shift in peak position to lower spatial frequency. The experimentally measured power spectra smoothly interpolate between the expectations for the 1D and 3D limits, with progressively closer agreement with the 1D stochastic model as the 1D regime is approached. Towards the 3D limit, the 1D theory \cite{Gillespie_1996, Stimming_2010, Imambekov_2009, Langen_2013, Schemmer_2018} becomes less valid, which manifests in the gradual suppression of the phase fluctuations as compared to the 1D theory extrapolated outside of its validity regime. Possible driving mechanisms include increased coherence due to the number of occupied transverse modes and/or the change from a phase coherence length to volume. Such details require further theoretical investigation, whether through calculation of higher order Bogoliubov modes or a stochastic Gross-Pitaevskii finite temperature approach \cite{Blakie_2008,Proukakis_2008_Finite}.

We extract the position of the first maximum $q_\mathrm{pk}$ and the peak amplitude $\langle|\rho(q_\mathrm{pk})|^2\rangle$. These are obtained by fitting the experimental spectra with the analytic result provided in \cite{Schemmer_2018} --- as we are only interested in matching the functional form of the spectra all fitting parameters ($T$, $N$, $t_\mathrm{tof}$, and the width of the imaging point spread function $\sigma_\mathrm{psf}$) are unconstrained. The relevance of the peak position is discussed in the Supplemental Material \cite{Imaging}. The average peak amplitude \mbox{$\langle|\rho(q_\mathrm{pk})|^2\rangle$} is a pertinent quantity to monitor in the crossover regime as it directly relates to the strength of the fluctuations in-trap--and thus, indirectly, the dimensionality of the system--while being only affected by interactions during time-of-flight expansion towards the 3D regime \cite{IntLessThan4}. While studying the peak amplitude as a function of $\tilde{\mu}$ and $\tilde{T}$, in addition, to put our work into context of earlier studies \cite{Dettmer_2001, Hellweg_2001, Menotti2002, Manz_2010, Perrin2012, Gallucci2012_PhaseyThings, Garret_2013}, we also monitor the dependency of \mbox{$\langle|\rho(q_\mathrm{pk})|^2\rangle$} on the trap aspect ratio \mbox{$\kappa = \omega_\perp/\omega_z$}, as well as the ratio of cloud length to the thermal phase coherence length \mbox{$\tilde{L} = L/\lambda_T$}, where \mbox{$\lambda_T = 2\hbar^2n(0)/mk_\mathrm{B}T$}. Note that $\tilde{L}$ is equivalent to ascribing a threshold temperature $T_\phi = \hbar^2n_0/mk_\mathrm{B}L$ \cite{Petrov_2001} below which phase fluctuations are suppressed, as has been used in previous literature \cite{Hugbart2005, Bouyer2003_TTphi, Gallucci2012_PhaseyThings}. The results are shown in Fig.~\ref{fig:Figure_2}.

Figure~\ref{fig:Figure_2}\textcolor{blue}{(a)} shows that at fixed $\kappa$ the size of the fluctuations can vary dramatically, i.e., the dimensionality is not solely driven by $\kappa$ in agreement with the current understanding that this is not a critical parameter. In a 1D gas at finite temperature, the two-point phase correlation decays exponentially with a characteristic length scale of $\lambda_T$. In practice, experiments explore systems with finite length $L$, so that $\tilde{L}$ becomes a quantity determining whether or not phase fluctuations are actually observed \cite{Petrov_2001, Hellweg_2003, Hugbart2005}. For \mbox{$\tilde{L}\lesssim 1$} a 1D system can have the appearance of a 3D system, but this can be interpreted as a finite size effect rather than a consequence of changed dimensionality. In an actual 3D gas, the quantity $\lambda_T$ loses its physical significance and phase fluctuations are strongly suppressed as long as $T$ is below the critical temperature for condensation. Figure~\ref{fig:Figure_2}\textcolor{blue}{(b)} shows an expected reduction of $\langle|\rho(q_\mathrm{pk})|^2\rangle$ as \mbox{$\tilde{L}$} decreases. At higher values of $\tilde{L}$, a spread of values for $\langle|\rho(q_\mathrm{pk})|^2\rangle$ occurs, supporting the notion that $\tilde{L}$ on its own is not indicative of the dimensionality, since our data show that chemically 1D clouds can have the same $\tilde{L}$ as clouds in the 1D to 3D crossover.

Figure~\ref{fig:Figure_2}\textcolor{blue}{(c)} shows the amplitude of the phase fluctuations over the $\tilde{\mu}$--$\tilde{T}$ parameter space explored. Here the chemical potential is determined from a local density approximation \mbox{$\tilde{\mu} = \big( \sqrt{1+4an(0)} - 1\big)$}. These data indicate that $\tilde{T}$ plays little role in determining the size of the fluctuations when the system is chemically 3D. For example, when \mbox{$\tilde{\mu} \gtrsim 4.5$} the fluctuations are always small despite the large variation in $\tilde{T}$ (interestingly, a recent numerical study of the 2D to 3D crossover found that effects of reduced dimensionality become observable below a similar threshold of \mbox{$\tilde{\mu} \sim 4$} \cite{Keepfer_2D23_2022}). However, when $\tilde{\mu}$ decreases the importance of $\tilde{T}$ on the fluctuations in the system increases, as can be seen in the inset of Fig.~\ref{fig:Figure_2}\textcolor{blue}{(c)}. These observations indicate that in the crossover regime the role of $\tilde{\mu}$ dominates that of $\tilde{T}$, quantitatively supporting previous qualitative statements in the literature 
\cite{Gorlitz2001_LowD_BECs, vanAmerongen2008_YYonAtomChip}. 

\begin{figure}[t!]
 \centering
\includegraphics[width=0.45\textwidth]{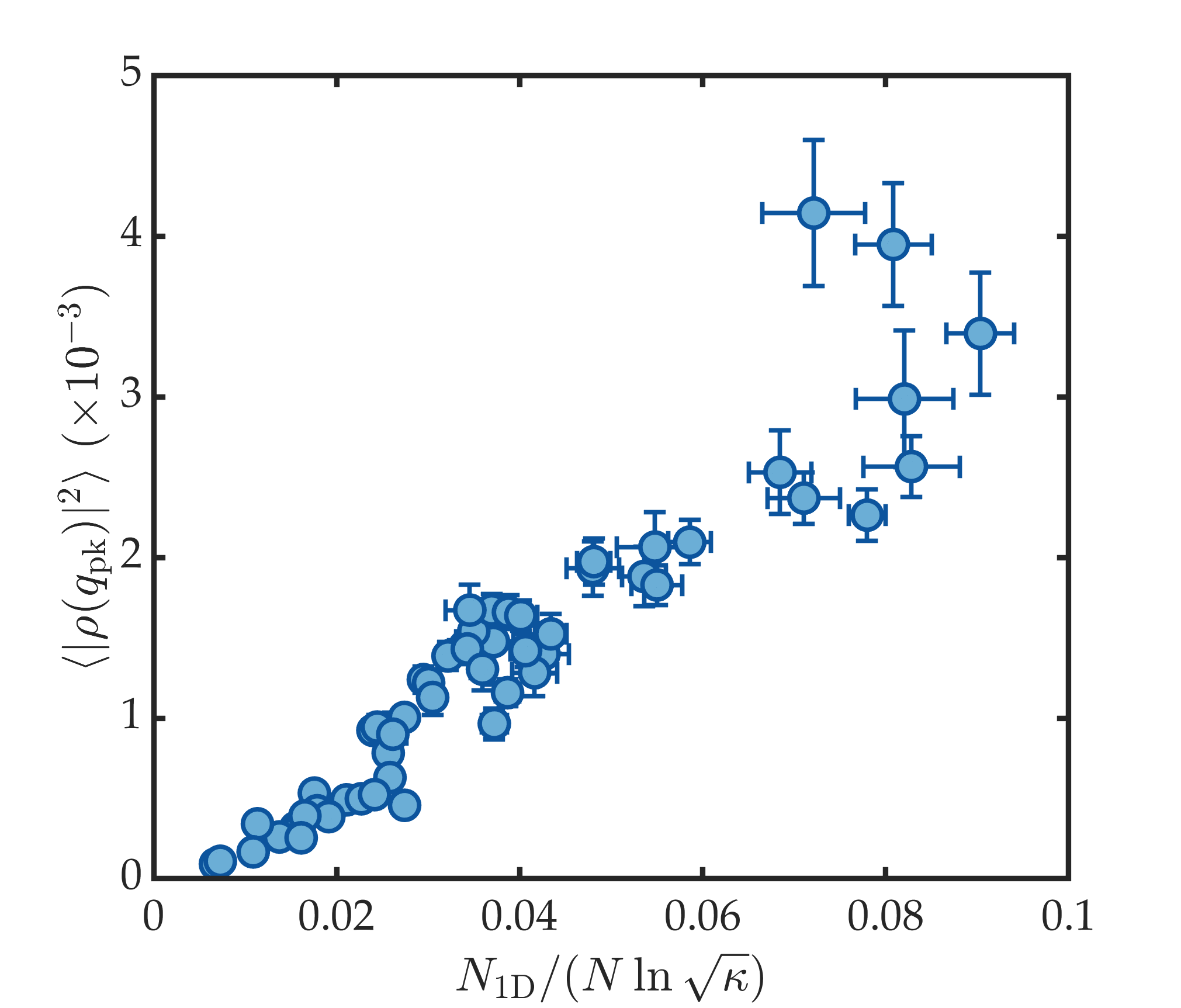}
\caption{\label{fig:Figure_3} Dependency of the amplitude of the density ripple power spectrum $\langle|\rho(q_\mathrm{pk})|^2\rangle$ on the relative number of 1D excitations $N_\mathrm{1D}/N$. Error bars for the data points represent the statistical uncertainty as two standard deviations from the mean, and are obtained by bootstrapping.}
\end{figure}


Since it is understood that fluctuations of the phase arise due to excitations in the system, we now examine the data in terms of the occupation of these modes, for predicting the strength of the phase fluctuations. Excitations of the quasi-condensate can be split into two categories --- \textit{high-energy} (\mbox{$\epsilon_j > \hbar\omega_\perp$}), and \textit{low-energy} (\mbox{$\epsilon_j < \hbar\omega_\perp$}) axial excitations. The latter exhibit a 1D character, with wavelengths larger than the radial size of the cloud but smaller than its axial size \cite{Petrov_2001}. Thus, only the low-energy excitations contribute to fluctuations of the phase. The axial spectrum of low-energy excitations for an elongated 3D condensate is \mbox{$\epsilon_j = \hbar\omega_z\sqrt{j(j+
3)/4}$} \cite{Stringari_1998}, and if \mbox{$k_\mathrm{B}T \gg \hbar\omega_z$} the occupation of each mode $j$ can be approximated to \mbox{$N_j = k_\mathrm{B}T/\epsilon_j$}. We propose that the relevant quantity is the relative population of 1D excitations present in the system, \mbox{$N_\mathrm{1D}/N$}, where
\begin{equation}
    N_\mathrm{1D} = \sum_j^{\epsilon_j < \hbar\omega_\perp} k_\mathrm{B}T/\epsilon_j.
    \label{Eq:N1D}
\end{equation}
This quantity compares the number of quasiparticles contributing to the phase fluctuations with the total number of atoms in the quasicondensate, which we expect to be related to the contrast of the observed density ripples and therefore $\langle|\rho(q_\mathrm{pk})|^2\rangle$. The relationship is shown in Fig.~\ref{fig:Figure_3}, where we observe a monotonic universal dependency. In contrast to when the data are plotted against $\kappa$ or $\tilde{L}$ (Fig.~\ref{fig:Figure_2}), it is striking that our data, which cover a broad experimental parameter space, now appear to collapse onto a single, approximately linear curve. This is a strong indication that \mbox{$N_\mathrm{1D}/N$} is the more relevant quantity for predicting the strength of the phase fluctuations.

In conclusion, we have performed a detailed experimental study of the onset of phase fluctuations in degenerate Bose gases in equilibrium throughout the full 1D to 3D dimensional crossover. We observe that the previously developed 1D stochastic model correctly describes the data only in one dimension, but exhibits a gradual departure from the experimental data as the dimensionality is tuned towards three dimensions. On the 3D side, an almost complete suppression of fluctuations is measured, as expected for a coherent 3D BEC. We confirm the expectation that $\tilde{\mu}$ determines whether the system exhibits 1D character. We find that for \mbox{$\tilde{\mu} \lesssim 4.5$} the gas displays phase fluctuations with a strength that is then also clearly dependent on $\tilde{T}$. In contrast, for larger values of $\tilde{\mu}$ the system appears effectively 3D regardless of $\tilde{T}$--as expected, since in three dimensions long-range order is possible even at finite temperature. The temperature dependence in the low-$\tilde{\mu}$ regime is understood in terms of the number of low-energy axial modes that can be populated, and that the fluctuations can indeed be measured by the relative population of these modes. Studies such as this which investigate the point at which a system passes through a dimensional crossover are of general interest--especially in other fields when particular regimes can be technically difficult to access. An important extension to this work will be to realize a similar experiment in a nonequilibrium setting with rapid changes from 1D to 3D and vice versa. It would also be of great interest to experimentally explore all possible crossovers involving any dimension, from 0D (where excitations are frozen along all directions) to 3D. Such a setup can be realized with cold atom systems by combining established techniques, including atom chips and optical-dipole traps. 

The authors would like to thank J. Dunningham for helpful discussions and careful reading of the manuscript. The work was financed by the University of Sussex Strategic Development Fund (SDF).

\bibliography{references}

\begin{thebibliography}{66}
\bibitem [{Int()}]{IntLessThan4}%
  \BibitemOpen
  \href@noop {} {}\bibinfo {note} {As can be seen in Fig.~\ref{fig:Figure_1}
  interactions have a minimal effect on the spectra of clouds with
  \mbox{$\tilde{\mu} \lesssim 4$}.}\BibitemShut {Stop}%
\end{thebibliography}

\end{document}


\preprint{APS/123-QED}

\title{Supplemental material for: Probing the Degree of Coherence through the Full 1D to 3D Crossover}

\author{R. Shah}
\author{T. J. Barrett}
\affiliation{Department of Physics and Astronomy, University of Sussex, Brighton BN1~9QH, United Kingdom}%
\author{A. Colcelli}
\affiliation{SISSA and INFN, Sezione di Trieste, Via Bonomea 265, I-34136 Trieste, Italy}
\author{F. Oru\v{c}evi\'{c}}
\affiliation{Department of Physics and Astronomy, University of Sussex, Brighton BN1~9QH, United Kingdom}
\author{A. Trombettoni}
\affiliation{SISSA and INFN, Sezione di Trieste, Via Bonomea 265, I-34136 Trieste, Italy}
\affiliation{Department of Physics, University of Trieste, Strada Costiera 11, I-34151 Trieste, Italy}
\author{P. Kr\"{u}ger}
\affiliation{Department of Physics and Astronomy, University of Sussex, Brighton BN1~9QH, United Kingdom}
\affiliation{Physikalisch-Technische Bundesanstalt, 10587 Berlin, Germany}

\date{\today}

\maketitle{}
\section{A. Interactions during time of flight} \label{appendix:Interactions}

In one-dimensional (1D) systems the in-trap interaction energy is small and is quenched quickly during time of flight (TOF) $t_\mathrm{tof}$, and it has been shown that in this case a purely ballistic expansion model is sufficient to describe the density ripples \cite{Imambekov_2009, Manz_2010}. However, for clouds in the 1D to 3D crossover the influence of the mean-field interaction during TOF is not negligible, and the expansion should described by the Gross-Pitaevskii equation. In the early stages of expansion ($t_\mathrm{tof} \lesssim 1/\omega_\perp$) the relative amount of interaction energy present is significant, and therefore the cloud undergoes hydrodynamic expansion. At longer times (\mbox{$t_\mathrm{tof} \gg 1/\omega_\perp$}) the density has reduced significantly and the interaction energy is effectively gone, and the cloud then evolves according to the linear Schr\"{o}dinger equation with the expansion being said to be ballistic. 

\begin{figure}[t]
\centering
\includegraphics[width=0.48\textwidth]{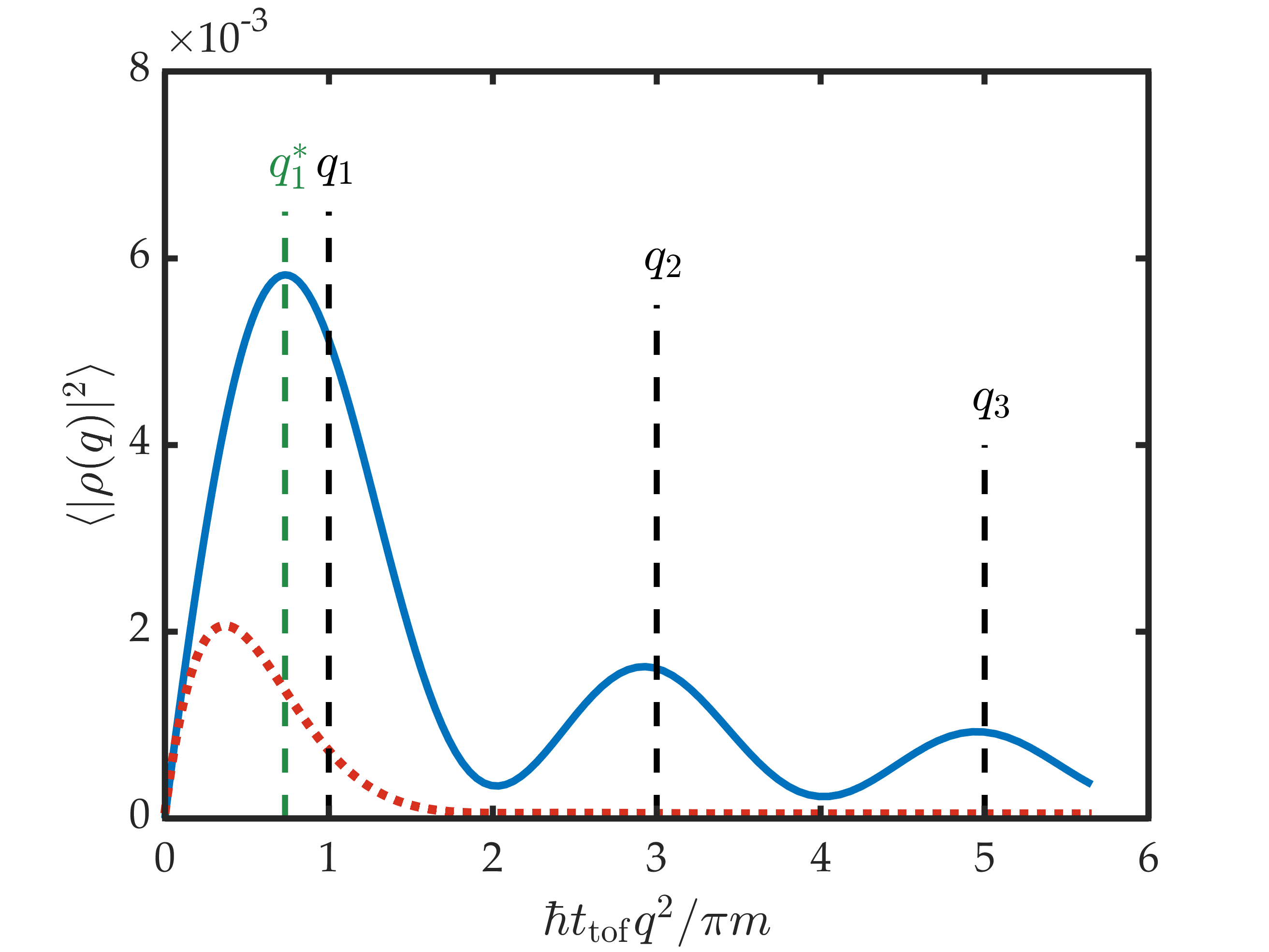}
\caption{\label{fig:PeakPositions} Peak positions in the power spectrum $\langle|\rho(q)|^2\rangle$, for an example system with trap frequencies $\omega_\perp / 2\pi=1$ kHz, $\omega_z / 2\pi=10$ Hz, and $10^4$ atoms at a temperature of $10$ nK. (Blue solid line) Analytical result from \cite{Imambekov_2009} which assumes pure ballistic expansion. (Red dotted line) The same result but including the effects of imaging with a finite point spread function by convolution with a Gaussian function with $\sigma=\SI{4}{\micro\meter}$. The peak predictions $q_n$ from Eq.~\ref{eq:q_n} are shown (black dashed line), along with the true position of the first peak $q_1^*$ (green dashed line). }
\end{figure}

In the case of ballistic expansion, the transverse and axial dimensions are separable, and the main contribution to the axial velocity distribution comes from the axial gradient of the in-trap phase \mbox{$\phi(z,t=0)$}, given by
\begin{equation}
    v(z,t) = \frac{\hbar}{m}\nabla\phi(z,t),
    \label{eq:velocityPhaseGradients}
\end{equation}
where $m$ is the atomic mass. After time of flight this gives rise to interference and the formation of density ripples. The phenomenon is the matter-wave analog of the temporal Talbot effect \cite{Talbot1836, Deng1999_TalbotEffect, Zhai2018_TalbotEffect, Santra2017_TalbotEffect, Chapman1995_TalbotEffect, Manfred2011_TalbotEffect, Makhalov2019_TalbotEffect}. Since in trap the phase effectively fluctuates at all spatial frequencies $q$, each particular $t_\mathrm{tof}$ will be equal to the Talbot time $t_\mathrm{Talbot}$ for some specific value of $q$, effectively amplifying the power spectrum $\langle|\rho(q)|^2\rangle$ at that $q$. In fact, fractional Talbot times also exist, i.e. \mbox{$t_\mathrm{Talbot}/n$ for $n = \{1,2,3 ...\}$}, and as such the power spectrum displays multiple discretely spaced peaks with decreasing amplitude - an example is shown in Fig.~\ref{fig:PeakPositions}. The $n^\mathrm{th}$ peak position $q_n$ is approximately given by \cite{Imambekov_2009},
\begin{equation}
    q_n = \sqrt{\frac{\pi m(2n-1)}{\hbar t_\mathrm{tof}}},
    \label{eq:q_n}
\end{equation}
and becomes more accurate for large $n$ (see Fig.~\ref{fig:PeakPositions}). Accounting for finite optical imaging resolution suppresses all but the first peak $q_1$ (to avoid confusion in the main text we refer to this peak $q_1$ as $q_\mathrm{pk}$). 

To account for the hydrodynamic expansion, we perform a numerical GPE calculation as follows. For a given atom number and trapping geometry we first numerically find the zero-temperature ground state of the system using the split-step Fourier method \cite{Taha_1984,Jackson_1998,Javanainen_2006,Antoine_2013} in imaginary time \cite{Dalfovo_1996,Bao_2004}. To model the effect of finite temperature we then imprint on this ground state wave function a phase that fluctuates axially but is invariant along the transverse direction. The phase is generated by a stochastic method which reproduces the correlation statistics from the Bogoliubov theory 
\begin{equation}
    \label{SM:eq1}
    \langle [\phi(z)-\phi(0) ]^2 \rangle = z/\lambda_T,
\end{equation}
where \mbox{$\lambda_T = 2\hbar^2n(0)/mk_\mathrm{B}T$} is the phase coherence length, and  works equally well in both the 1D \cite{Stimming_2010} and elongated 3D regimes \cite{Petrov_2001}.

\begin{figure}[h]
\centering
\includegraphics[width=0.48\textwidth]{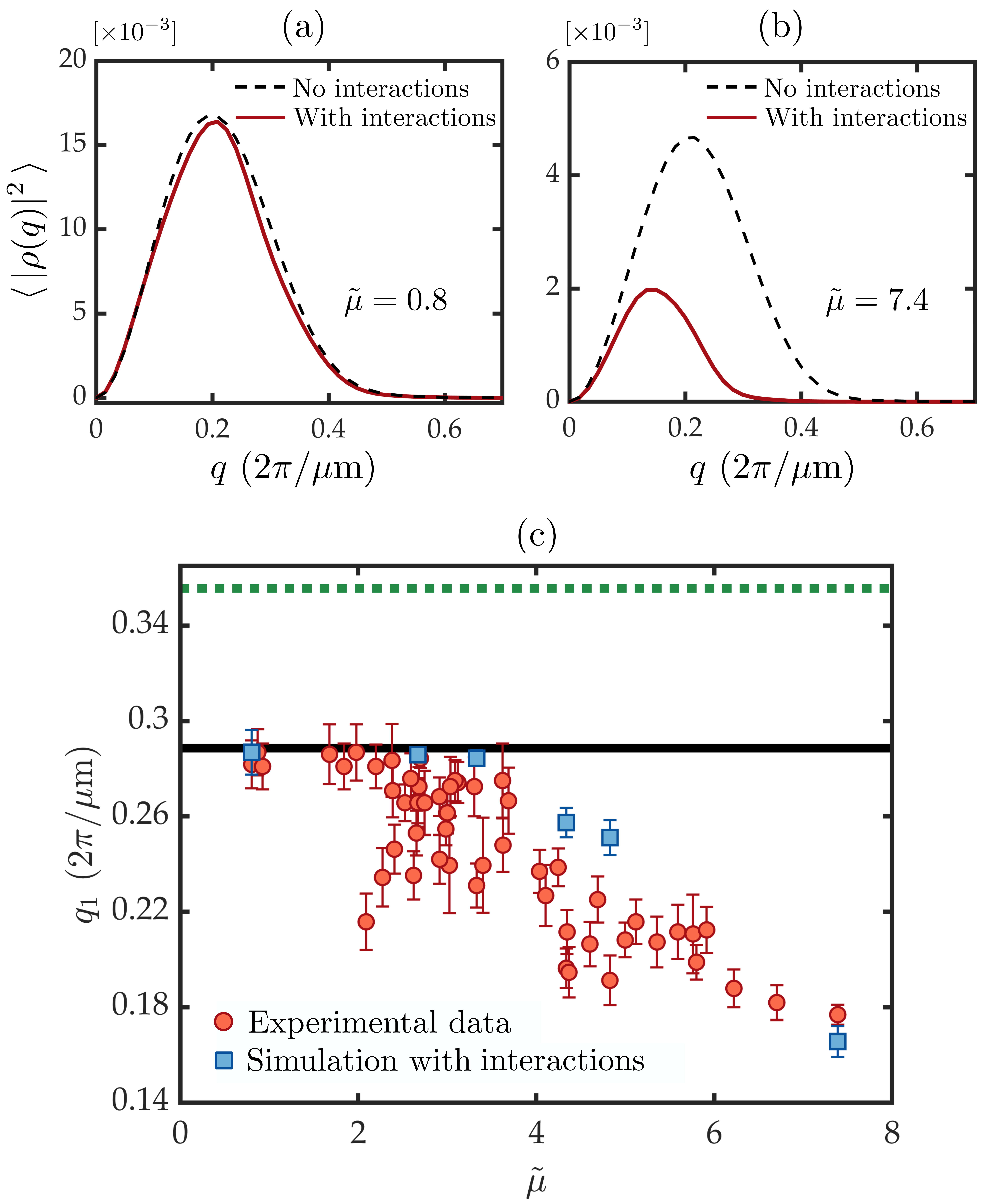}
\caption{\label{fig:InteractionsOnPowerSpectrum} A comparison of simulated power spectra for clouds undergoing hydrodynamic and ballistic expansion, for a 1D quasicondensate (a) and an elongated 3D BEC (b). (c) Effect of interactions on the first peak position $q_1$ of the power spectra. The green dotted line shows $q_1$ calculated with Eq.~\ref{eq:q_n} for a time of flight of $t_\mathrm{tof} = 34$ ms. Experimental data  (red circles) and simulated data including interactions (blue squares) for comparison with Eq.~\ref{eq:q_n} the effect of the imaging system has been removed. (Black solid line) Mean peak position extracted from simulated spectra excluding interactions, labelled as $q_1^*$ in Fig.~\ref{fig:PeakPositions}. Vertical error bars show two standard deviations from the mean and are estimated by bootstrapping.}
\end{figure}

Next, to calculate the time-of-flight expansion the wave function with fluctuating phase is evolved in real time using the GPE with the trapping potential removed until the interaction energy becomes negligible (typically \mbox{1 -- 6 ms}, depending on the transverse trapping frequency). We then further expand this wave function ballistically (ignoring interactions) using the free Schrodinger equation out to the final time of flight of \SI{34}{\milli\second} \cite{Deuar2016}. The process is repeated with 200 realisations of random phases, and the resulting simulated images are then analyzed in same way as the experimental data to obtain the power spectrum of the density ripples $\langle|\rho(q)|^2\rangle$. Examples of power spectra for clouds in the 1D and 3D regime, with \mbox{$\tilde{\mu} = 0.8$ \textrm{and} $7.4$} respectively, are shown in Fig.~\ref{fig:InteractionsOnPowerSpectrum}\textcolor{blue}{a--b}. In this figure, we also compare the results of the calculation with a purely ballistic expansion model (this time neglecting interactions throughout the entire time-of-flight evolution) to directly reveal the effects of interactions. As is apparent in the 3D case the interactions not only suppress the peak amplitude of the power spectrum but also shift it to a lower spatial frequency. In contrast, the influence of the interactions in the 1D regime is almost negligible. We note that Ref. \cite{Dettmer_2001} provides an analytic treatment for calculating the density ripples including interactions during time of flight, making use of the hydrodynamic scaling solution \cite{Castin_1996}. However, after comparison with our GPE results, we found that the approach breaks down when the axial expansion becomes significant (as is the case with the tighter axial trapping frequencies used here).

The effect can be understood as follows, during the period of hydrodynamic expansion repulsive interactions between atoms induce an acceleration, broadening the velocity distribution, and cause the density ripples to spread out further than they would if the expansion was ballistic. This interaction induced spreading of the density ripples effectively suppresses the power spectrum not only reducing the amplitude but also shifting the spectrum to lower spatial frequencies, as can be seen in Fig.~\ref{fig:InteractionsOnPowerSpectrum}\textcolor{blue}{b}. To observe this effect we extract the position of the first maximum $q_1$ from all of the experimental data, with the results shown in Fig.~\ref{fig:InteractionsOnPowerSpectrum}\textcolor{blue}{c}. We observe a clear correlation in peak position with the reduced chemical potential $\tilde{\mu}$ through the 1D to 3D crossover and find clear agreement with our GPE simulation. Thus, the significance of interactions increases as the dimensionality gradually changes from one dimension to three dimensions and should be accounted for when modelling systems within the 1D to 3D crossover. However, they can be safely ignored close to or inside the 1D regime. The same shifting of the spectral peak positions was predicted and observed in 2D degenerate Bose gases \cite{Mazets_2012, Seo_2014}, however there was a small disparity between the prediction and observation.

\section{B. Imaging} \label{appendix:Imaging}

\begin{figure}[t]
\includegraphics[width=0.49\textwidth]{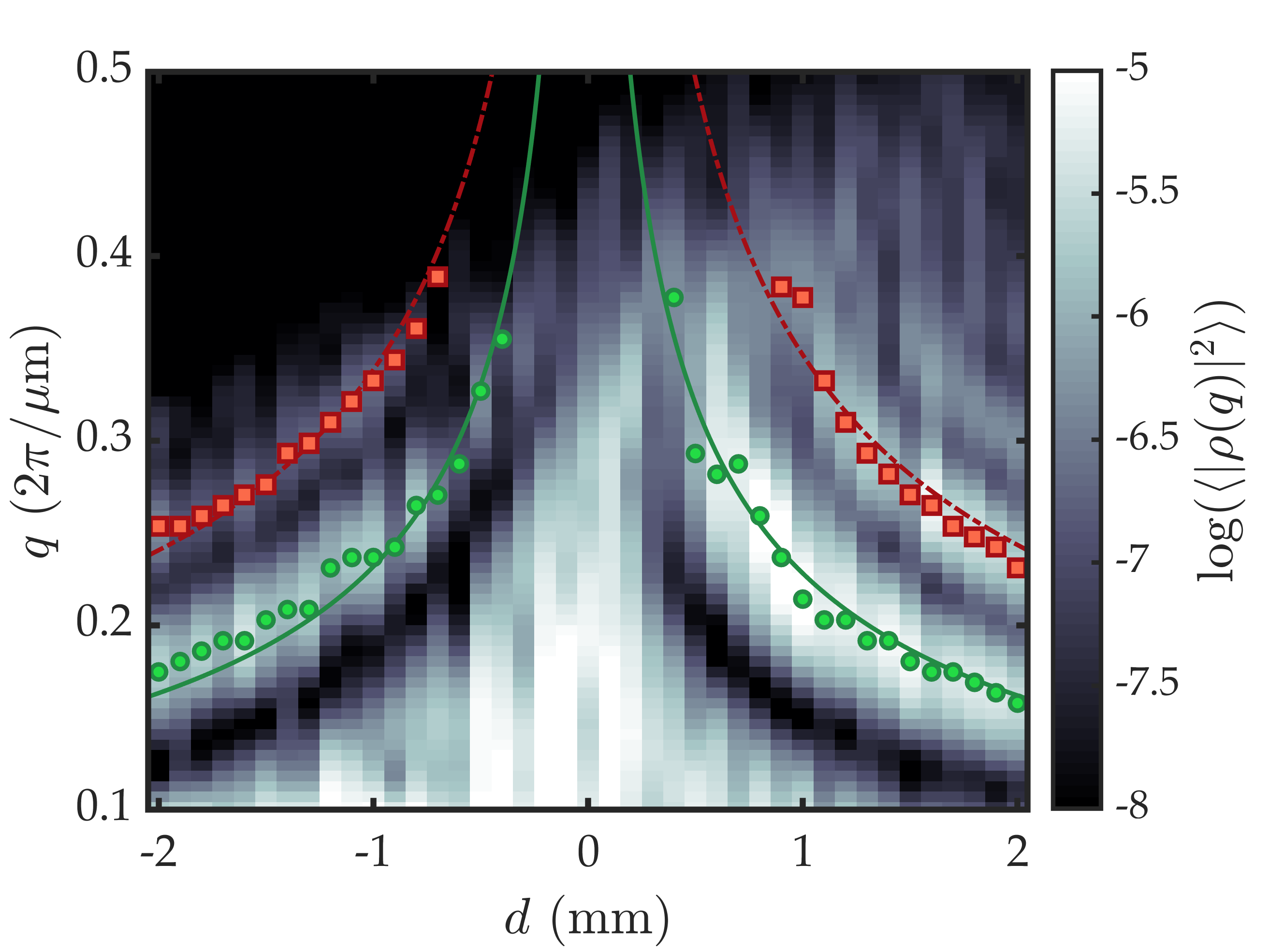}
\caption{\label{fig:A1_focus} Power spectrum of the density ripples $\langle|\rho(q)|^2\rangle$ across the focus. At each position $d$ the power spectrum $\langle|\rho(q)|^2\rangle$ is calculated from approximately 30 images. Green (Red) circles (squares) are the measured second (third) peak positions, and the green (red) solid (dash-dotted) line is a fit using Eq.~\ref{eq:A1}. The optimal focus position is indicated by $d = 0$ mm with an uncertainty of $\pm$ 20 $\upmu$m.}
\end{figure}

To focus our optical imaging system on the plane of the atomic cloud we have used the technique described in \cite{Putra_2014}. This method utilizes the sensitivity of the power spectrum $\langle|\rho(q)|^2\rangle$ to defocusing effects which can be detrimental to measurements of phase fluctuations. As the imaging system is moved out of focus by a distance $d$ from the focal plane the density ripples are blurred, and additional fringes appear since it is now rather the near-field diffraction pattern which will be imaged onto the camera. In the power spectrum this results in an attenuation of the amplitude and the creation of additional higher frequency maxima. The modified power spectrum $\langle|\rho(q,d)|^2\rangle$ is related to the in-focus power spectrum $\langle|\rho(q,d=0)|^2\rangle$ by
\begin{equation}
\label{eq:A1}
\langle|\rho(q,d)|^2\rangle =  \langle|\rho(q,0)|^2\rangle \cos^2 \left( \frac{q^2 d}{2 k_0} \right),
\end{equation}
where $k_0$ is wave number of the probe light. A measurement of $\langle|\rho(q,d)|^2\rangle$ is shown in Fig.~\ref{fig:A1_focus}, where the second and third order peaks (arising from the diffraction fringes) can be seen clearly as the system is moved out of focus. The optimal focal position $d=0$ is determined by fitting the higher order peaks with Eq.~\ref{eq:A1}.

Finite optical resolution can be accounted for in a simple approximation by convolving the simulated density ripples with the point spread function (PSF) of the imaging system. The PSF of a diffraction-limited imaging system has the functional form of an Airy disk, which we approximate by fitting a Gaussian to the central lobe. Convolution of the density ripples with the Gaussian-approximated PSF modifies the experimentally detected power spectrum in \mbox{$q$-space}, which is then given by
\begin{equation}
 \langle|\rho(q)|^2\rangle_\mathrm{exp} =  \langle|\rho(q)|^2\rangle \, e^{-\sigma_\mathrm{psf}^2q^2},
    \label{eq:A2}
\end{equation}
where $\sigma_\mathrm{psf}$ is the RMS width of the Gaussian. To obtain the value of $\sigma_\mathrm{psf}$ we fitted the measured power spectrum of clouds in the 1D limit $\tilde{\mu}\lesssim 1$ where the effect of interactions during expansion are negligible, yielding a value of $\sigma_\mathrm{psf} \sim 4$ $\upmu \mathrm{m}$.

\section{C. Power spectrum scaling}\label{appendix:Scaling}

In our analysis we require a quantity that is independent of atom number and the size of the system. The Fourier transform of the density ripples given by Eq. \textcolor{blue}{1} has effective units of atom number, and thus to be truly dimensionless $|\delta\tilde{n}(q)|^2$ must be scaled by $1/N^2$. To understand the use of Eq. \textcolor{blue}{2} to investigate 1D character, we consider here a simple case in the 1D limit. Starting with the approximation for a homogeneous gas including only small wave vectors \mbox{$q\hbar t_\mathrm{tof}/m \ll \lambda_T$} the power spectrum of density ripples is given by \cite{Schemmer_2018}
\begin{equation}
    \frac{\langle|\rho_{n_0}(q)|^2\rangle}{N^2} = \frac{1}{N^2} 4 n_0^2 \langle \theta_q^2 \rangle \sin^2\left[ \hbar q^2 t_\mathrm{tof} / 2m \right],
    \label{eq:HomoPSD_low-q}
\end{equation}
where \mbox{$\langle\theta_q^2\rangle = mk_BT/\hbar^2n_0q^2 $} is the phase quadrature describing the in-trap phase fluctuations. To account for inhomogenous densities, we can perform a local density approximation using the homogeneous result of Eq. \ref{eq:HomoPSD_low-q} and introducing a spatial dependence on the density $n_0(z)$
\begin{equation}
    \frac{\langle|\rho(q)|^2\rangle}{N^2} = \frac{1}{N^2} \int_{-R}^{R}  \langle | \rho_{n_0(z)}(q)|^2\rangle \,dz,
    \label{eq:LDA}
\end{equation}
where \mbox{$ R=\sqrt{2\mu/m\omega_z^2} $} is the Thomas-Fermi radius. Combining equations \ref{eq:HomoPSD_low-q} and \ref{eq:LDA}, we obtain
\begin{equation}
    \frac{\langle|\rho(q)|^2\rangle}{N^2} = \frac{1}{N^2} \int_{-R}^{R}  \frac{4mk_BTn(z)}{\hbar^2q^2} \sin^2\left[ \hbar q^2 t_\mathrm{tof} / 2m \right] \, dz.
    \label{eq:LDA_low-q}
\end{equation}
Using the density profile of a pure 1D Thomas-Fermi condensate \mbox{$n(z) = n_0\left[ 1 - \left( z/R\right)^2 \right]$}, and evaluating the integral, we arrive at the final expression
\begin{equation}
    \frac{\langle|\rho(q)|^2\rangle}{N^2} = \frac{16mk_BTn_0R}{3\hbar^2q^2 N^2} \sin^2\left[ \hbar q^2 t_\mathrm{tof} / 2m \right].
    \label{eq:Low-q_PSD}
\end{equation}
Substituting for $R$ and $N=4n_0R/3$ the power spectrum scales with $\omega_z$ (i.e. the length of the system)
\begin{equation}
    \frac{\langle|\rho(q)|^2\rangle}{N^2} = \omega_z \sqrt{\frac{m}{2\mu}} \frac{3mk_BT}{n_0 \hbar^2 q^2}\sin^2\left[ \hbar q^2 t_\mathrm{tof} / 2m \right]
    \label{eq:PSD_N2}.
\end{equation}
Since the axial trap frequency is varied in our experiment, we divide this quantity by $\tau = \omega_z t_\mathrm{tof}$ to remove the dependency on the length of the system and keep the power spectrum dimensionless, arriving at
\begin{equation}
    \frac{\langle|\rho(q)|^2\rangle}{\tau N^2} =  \sqrt{\frac{m}{2\mu}} \frac{3mk_BT}{n_0 \hbar^2 q^2 t_\mathrm{tof}  }\sin^2\left[ \hbar q^2 t_\mathrm{tof} / 2m \right]
    \label{eq:PSD_wtN2}.
\end{equation}
In our study the time of flight is fixed across the entire data set and hence does not effect the physics. This final expression is now appropriately normalized.

\bibliography{supplementaryRefs}